\documentclass[aps,prb,reprint,superscriptaddress,showpacs,amsmath,amssymb]{revtex4-1}
\usepackage{epsfig,psfrag}
\usepackage{dcolumn}
\usepackage{bm}
\usepackage{graphicx}
\usepackage{color}
\usepackage{braket}

\begin{document}
\title{One dimensional massless Dirac bands in semiconductor superlattices}

\author{F.Carosella}
\email{Electronic address: francesca.carosella@lpa.ens.fr}
\affiliation{Laboratoire Pierre Aigrain, ENS-CNRS UMR 8551, Universit\'es P. et M. Curie and Paris-Diderot,
24, rue Lhomond, 75231 Paris Cedex 05, France} 

\author{A. Wacker}
\affiliation{Mathematical Physics, Lund University, Box 118, S-22100 Lund, Sweden} 

\author{R. Ferreira}
\affiliation{Laboratoire Pierre Aigrain, ENS-CNRS UMR 8551, Universit\'es P. et M. Curie and Paris-Diderot,
24, rue Lhomond, 75231 Paris Cedex 05, France} 

\author{G. Bastard}
\affiliation{Laboratoire Pierre Aigrain, ENS-CNRS UMR 8551, Universit\'es P. et M. Curie and Paris-Diderot,
24, rue Lhomond, 75231 Paris Cedex 05, France}

\begin{abstract} 
Semiconductor superlattices may display dispersions that are degenerate either at the zone center or zone boundary \citep{Bastard1988wave,Sirtori1994}. We show that they are linear upon the wave-vector in the vicinity of the crossing point. This establishes a realisation of massless Dirac bands within semiconductor materials. We show that the eigenstates and the corresponding Wannier functions of these superlattices have peculiar symmetry properties. We discuss the stability of the properties of such superlattices versus the electron in-plane motion. As a distinct fingerprint, the inter-subband magneto-absorption spectrum for such superlattices is discussed.

\end{abstract}

\pacs{73.21.Ac,78.67.Pt}

\maketitle

Massless Dirac bands, electronic dispersion relations that are linear upon the
wave-vector in the vicinity of a high symmetry point in the Brillouin zone, are
heavily searched because they lead to unusual physical properties
\cite{Castro-Neto2009, Orlita2010}. The prototype of material that displays
such linear dispersion relations is graphene. Here, we will show that the very
same linear dispersions occur for the unbound states of one dimensional
semiconductor superlattices \cite{Esaki1970} (SL), like GaAs/Ga(Al)As,
provided the layer thicknesses are appropriately chosen. In the following we
will refer to such specific superlattices as Dirac L's.  The existence of
gap-less states (sub band crossing) in the continuum of superlattices was
briefly mentioned in \cite{Bastard1988wave} and their experimental evidence
was first obtained by Sir-tori et AC. \cite{Sirtori1994}  by means of
intersubband absorption in (CA,In)As/(Al,In)As superlattices. Going
beyond these studies we address the following issues here: (i) The
occurrence of a linear dispersion with an associated Dirac point is discussed.
(ii) The  change in the parity property of the SL eigenstates and of their
associated Wannier functions when crossing the Dirac point by 
a small change in the thicknesses is established.
(iii) The stability of the Dirac point with respect to the electron 
in-plane motion, which is non-trivial as the longitudinal and in-plane motions
are coupled in general. Furthermore we present our results for the 
inter-subband magneto-absorption where the Dirac SL's should be best evidenced.

\section{Analytical relation for a \protect\\ Dirac point}

\begin{figure}
\includegraphics[scale=0.4]{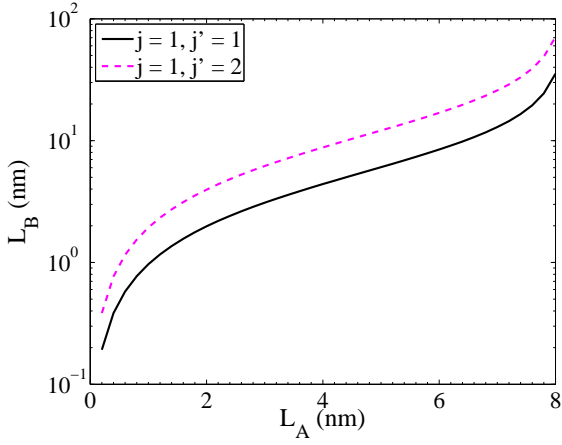}
\caption{\label{Fig.1}Barrier ($L_B$) versus well ($L_A$) thickness for the $j = j' = 1$ resonance condition (dashed line) and for the $j = 1$, $j' = 2$ resonance condition (continuous line).}
\end{figure}

Within the present work we consider a binary SL made of a periodic stacking of layers \textit{A} (well-acting material) and \textit{B} (barrier-acting material) with thicknesses $L_A$, $L_B$. We denote $d = L_A+L_B$ as
the SL period. We use parabolic dispersion relations in both kinds of layers characterized by effective masses $m_A$, $m_B$ in the well and barrier, respectively \cite{Bastard1981, *Bastard1982} (including band non parabolicity is doable if requested but cumbersome and does not bring any new feature to the linear dispersion problem \footnote{In the case of InAs/GaSb superlattices where the non parabolicity is a mandatory ingredient to understand the hybridization between the InAs electrons states and the GaSb light hole states, all the results discussed in terms of Fabry-Perot conditions as well as of linear dispersions apply.  The only ingredient that changes in eq.\ref{eq.1} is the definition of the effective masses $m_A$ and $m_B$ which must account for non parabolicity effects \cite{Bastard1981, *Bastard1982}.}). We choose the energy origin at the bottom of the conduction band of the well-acting material and call $V_b$ the barrier height. We note \textit{q} the SL wavevector and concentrate on the electron motion along the growth axis for states that are propagating in both kinds of layers ($\epsilon \geq V_b$). The superlattice dispersion relation for a zero in-plane wavevector, i.e. at the subband edge, is therefore \cite{Bastard1981, *Bastard1982}:
%
\begin{eqnarray}
\cos (qd)=\cos (k_{A}L_{A})\cos (k_{B}L_{B})-
\label{eq.1}
\end{eqnarray}
\begin{eqnarray}
-\frac{1}{2}(\xi+\frac{1}{\xi})\sin (k_{A}L_{A})\sin (k_{B}L_{B})
\nonumber
\end{eqnarray}
where
\begin{eqnarray}
\xi=\frac{k_{A}m_{B}}{m_{A}k_{B}}
\text{,}
\qquad
k_{A}=\sqrt{\frac{2m_{A}\epsilon}{\hbar^{2}}}
\text{,}
\qquad
k_{B}=\sqrt{\frac{2m_{B}(\epsilon -V_{b})}{\hbar^{2}}}
\nonumber
\end{eqnarray}
In eq.\ref{eq.1} one sees immediately that the energies $\epsilon_{jj'}$  which fulfill:
%
\begin{equation}
k_{A}L_{A}=j\pi
\qquad
k_{B}L_{B}=j'\pi
\label{eq.2}
\end{equation}
with $j$ and $j'$ integers are solutions of the equation.  If $j+j’'$ is even (odd) these energies are associated with $qd = 0$ ($qd = \pi$). This double Fabry-Perot condition was mentioned to be associated with zero bandgap in the SL dispersion relations \citep{Bastard1988wave,Sirtori1994,Bastard1991_SSP}. It implies a definite relationship between $L_A$, $L_B$ and $V_b$:
%
\begin{equation}
\frac{m_{B}j^{2}\pi^{2}}{m_{A}L_{A}^{2}}-\frac{2m_{B}V_{b}}{\hbar^{2}}=\frac{j'^{2}\pi^{2}}{L_{B}^{2}}
\label{eq.3}
\end{equation}
Hence, for masses that are not too different, $L_B$ has to be larger than $L_A$ if $j = j'$. We show in fig. \ref{Fig.1} the $L_B$ versus $L_A$ curve for $j = j' = 1$, and $j = 1$ and $j' = 2$ using the material parameters $m_A = 0.07m_0$, $m_B = 0.076m_0$, $V_b = 80$ meV. 
%
These parameters correspond roughly to GaAs/Ga$_{0.89}$Al$_{0.11}$As SL's. A low barrier height will ensure the Dirac bands to be easily optically probed and affect significantly the carrier dynamics in the SL. In this work we will show the results of the calculations for two Dirac SL's structures with the parameters indicated above and either with $L_A=7 $ nm and $L_B = 12.92 $ nm (satisfying the resonance condition $j = j'= 1$), or with $L_A = 6.6 $ nm and $L_B = 21.37 $ nm (satisfying the resonance condition $j = 1$, $j'=2$). 

The very fact that both sines in eq.\ref{eq.1} vanish when the resonance conditions (eq.\ref{eq.2}) are satisfied implies that close to an energy $\epsilon_{jj'}$  the dispersion relations will be linear either in the vicinity of $q = 0$ or $q = \pi/d$ and degenerate in one of these points. In fact letting $\epsilon=\epsilon_{jj'}+\eta$, with $\eta$ very small, for $j+j'$ odd and $q=\pi/d -Q$, with \textit{Qd} small and positive, there is:
%
\begin{eqnarray}
\eta^{2}=\frac{Q^{2}d^{2}}{G_{jj'}}
\label{eq.4}
\end{eqnarray}
\begin{multline*}
G_{jj'}=\frac{m_{B}^{2}L_{B}^{4}}{\hbar ^{4}j'^{2}\pi ^{2}}+\frac{m_{A}^{2}L_{A}^{4}}{\hbar ^{4}j^{2}\pi ^{2}}\\
+\left(\frac{jm_{B}L_{B}}{j'm_{A}L_{A}}+\frac{j'm_{A}L_{A}}{jm_{B}L_{B}}\right)
\frac{m_{A}m_{B}L_{A}^{2}L_{B}^{2}}{\hbar ^{4}jj'\pi ^{2}}
\end{multline*}
On the other hand for $j+j'$ even, in the vicinity of $q = 0$, we find a similar formula:
%
\begin{equation}
\eta^{2}=\frac{q^{2}d^{2}}{G_{jj'}}
\label{eq.5}
\end{equation}
where $G_{jj'}$ is the same as in eq.\ref{eq.4}. Hence, in contrast to a widespread belief, the dispersion relations of binary SL's can be linear in \textit{q} in the vicinity of either the Brillouin zone center or the zone boundary provided the double Fabry Perot conditions are fulfilled. The effective velocity corresponding to this linear dispersion close to $qd = \pi$ is $4.6 \times 10^{5} m/s$ for the Dirac SL with $j = 1, j' = 2$ resonance. This average velocity for a $\ket{p,q}$ SL state is equal to $\bra{p,q}\frac{p_z}{m_0}\ket{p,q}$ and coincides numerically with $\frac{1}{\hbar}\frac{\partial\epsilon_p}{\partial q}$ in spite of the inapplicability of the usual one band approximation to this degenerate case.

Note that for an arbitrary superlattice it is known \cite{Ashcroft1976} that the dispersion relations are the solution of the following equation
%
\begin{equation}
\cos (qd)=f(\epsilon)
\label{eq.6}
\end{equation}
where $f(\epsilon)$ is a function of the energy. Hence, to get Dirac bands in an arbitrary superlattice, the function $f(\epsilon)$ must be such that in the vicinity of  $\epsilon_c= \epsilon(q=0)$ or $\epsilon_b= \epsilon(q=\pi/d)$ there is:
%
\begin{equation}
f(\epsilon)\approx 1-\frac{(\epsilon-\epsilon_c)^2}{\delta _{c}^{2}}
\label{eq.7}
\qquad
\text{or}
\qquad
f(\epsilon)\approx -1+\frac{(\epsilon-\epsilon_b)^2}{\delta _{b}^{2}}
\end{equation}
where $\delta_c$ and $\delta_b$ are constants. It is difficult to be more specific on general grounds since $f(\epsilon)$  is fixed by the potential profile in the superlattice unit cell.  However, we note that the function $f(\epsilon)$ is usually larger or much larger than one when the electron wave is evanescent, thereby preventing eq.\ref{eq.7} to be realized.  
In addition, we wish to point out that the existence of Dirac bands in a given superlattice family (that differs by the strength of the potential or by the period length as found e.g. in the cosine-shaped potential $V(z)=V_{b}\cos (\frac{2\pi z}{d})$) is by no means guaranteed.  Let us indeed consider the Dirac comb:
%
\begin{equation}
V(z)=V_{0}L\sum_{n} \delta(z-nd)
\label{eq.8}
\end{equation}
where $L$ is a length and $d$ the period.  It is easily found that :
%
\begin{eqnarray}
\cos(qd)=f(\epsilon)=\cos(kd)+\frac{m^{*}V_{b}L}{\hbar^{2}}\frac{\sin(kd)}{kd}
\label{eq.9}
\end{eqnarray}
\begin{eqnarray}
k=\sqrt{\frac{2m^{*}\epsilon}{\hbar^{2}}}
\nonumber
\end{eqnarray}
It is still true that $kd = m\pi$, with $m$ an integer, ensures $f(\epsilon_{m})=(-1)^{m}$. However, at these energies it is impossible to simultaneously ensure $\frac{df}{d\epsilon}(\epsilon_m)=0$. Hence, in general, a one-dimensional potential does not always admit Dirac bands. For that reason in the present article we study only the specific case of flat band binary superlattices.

\begin{figure*}
\includegraphics[scale=0.9]{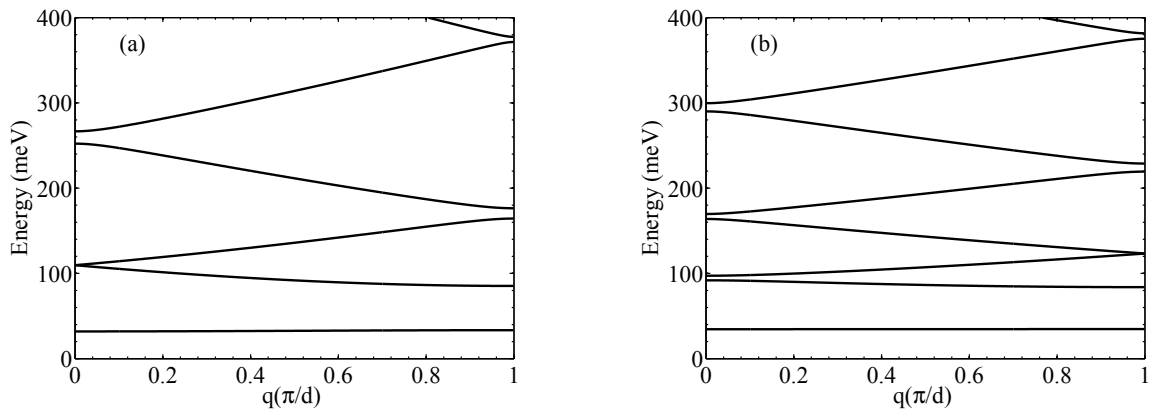}
\caption{\label{Fig.2}Dispersion relations for a GaAs/Ga(Al)As SL verifying the resonance condition (a)  $j = j' = 1$  ($V_b$ = 80 meV, $L_A$ = 7 nm, $L_B$ = 12.92 nm) and (b) $j = 1, j' = 2$  ($V_b$ = 80 meV, $L_A$ = 6.6nm, $L_B$ = 21.37 nm). Notice that in both cases the first subband is bound and is almost dispersionless. In panel (a) the 2\textsuperscript{nd} and 3\textsuperscript{rd} subbands are Dirac-like, while in panel (b) the 3\textsuperscript{rd} and 4\textsuperscript{th} subbands are Dirac-like.}
\end{figure*}

We show in fig.\ref{Fig.2} the flat band binary SL dispersion relations $\epsilon_p(q)$ calculated for a $j = j' = 1$ resonance and for a $j = 1$, $j' = 2$ resonance (the parameters for each structure are indicated above). 
%
As expected from the analytical calculation (eq.\ref{eq.4} and eq.\ref{eq.5}) we find subbands with linear dispersions and degenerate at $q = 0$ or $q = \pi/d$. Specifically, for the SL with $j = j' = 1$ resonance the 2\textsuperscript{nd} and 3\textsuperscript{rd} subbands are degenerate at $q = 0$ and show linear dispersions close to the zone center. Conversely, for the SL with $j = 1$, $j' = 2$ resonance the 3\textsuperscript{rd} and 4\textsuperscript{th} subbands are degenerate at $q = \pi/d$ and are Dirac-like close to the zone boundary. In both cases there is a single subband bound in the well that exhibits very little dispersion (less than 1 meV).  The other subbands are regular SL subbands. 

\section{Wannier functions at the \protect\\ Dirac point}
\begin{figure}
\includegraphics[scale=0.4]{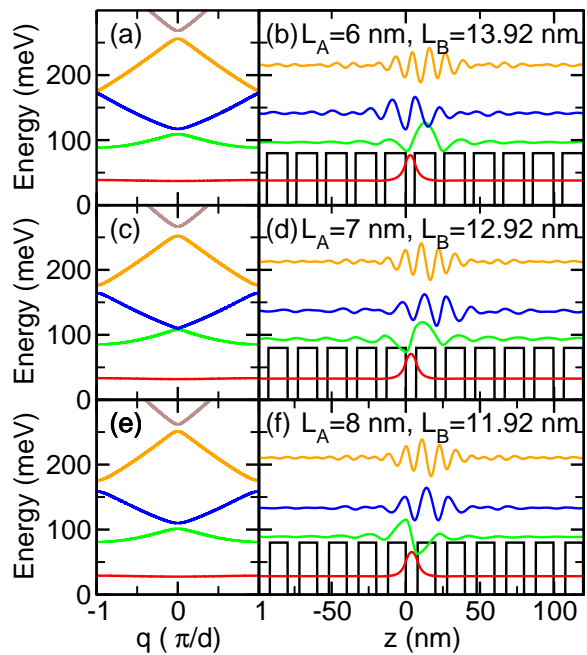}
\caption{\label{Fig.3} Wannier functions calculated according to the procedure of Ref. \onlinecite{Bruno-Alfonso2007} for a sequence of superlattices, where the middle one with $L_A$ = 7 nm, $L_B$ = 12.92 nm satisfies the Dirac condition.}
\end{figure}

Moreover, the realization of a resonance condition in a SL influences dramatically the symmetry properties of the SL eigenstates and of their associated Wannier functions. Wannier functions can be constructed from the 
Bloch states, for SL's see, e.g., Refs.~\onlinecite{Kohn1959_p1,Bruno-Alfonso2007}, where
the optimization of their spatial localization was addressed. We show in fig.\ref{Fig.3} a comparison between the Wannier functions of a Dirac SL ($j = j' = 1$) and those of SL's with nearby layer thicknesses, the SL period being kept the same. 
%
On general grounds \cite{Kohn1959_p1,Bruno-Alfonso2007}, the Wannier functions for a superlattice with inversion symmetry should be symmetric or antisymmetric with respect to one of the symmetry points (center of well or center of barrier). While the Wannier function of the bound subband is about the same in the three SL's, being symmetrical with respect to the center of the well, the symmetry property of subbands with energy larger than $V_b$ are interchanged in the sequence of SL's. In the case of the wider well ($L_A = 8$ nm), the Wannier function of the second subband is antisymmetric with respect to the center of the well, and the Wannier function of the third subband is symmetric with respect to the center of the barrier. Reducing the well width (and increasing the barrier width) increases the energy of the well-like state and decreases the energy of the barrier-like state, so that the sequence is opposite at $L_A = 6.5$ nm . In between (for the Dirac SL at $L_A = 7$ nm) the symmetries of these Wannier functions are becoming ill-defined. Furthermore, the Wannier functions for the Dirac SL (evaluated by the procedure of Ref.\onlinecite{Bruno-Alfonso2007}) are badly localized and we cannot observe an exponential decay numerically. 

\section{Absorption spectrum}

Linear dispersions imply a number of distinctive features.  For instance, the inter-subband absorption lineshape will be drastically modified compared to the usual divergences at subband extrema $q = 0$ or $q =\pi/d$ expected for a 1D free particle with quadratic dispersion relation \cite{Helm1993}. In the following, we discuss the intersubband absorption starting from the ground subband of the superlattice. We assume a strong magnetic field has been applied parallel to the growth axis in order to Landau quantize the in-plane motion ($\omega_c \approx 16.4$ meV at $B = 10$ T for GaAs). Under such circumstances, the electronic motion is free only along the growth axis.  The optical selection rules are that the electric vector of the wave has to be parallel to the growth axis and that the Landau quantum numbers are conserved for the in-plane motion and that the transitions are vertical in the reciprocal space. 

\begin{figure}
\includegraphics[scale=0.4]{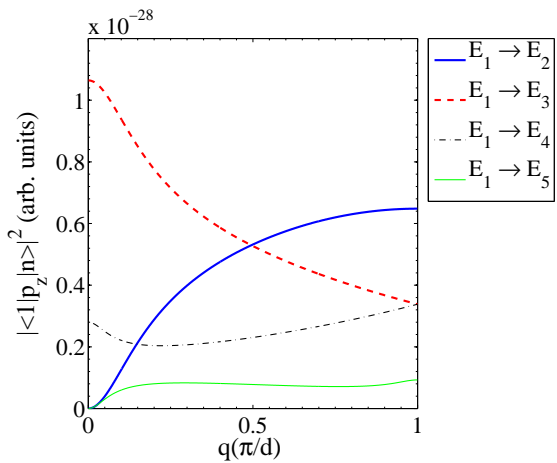}
\caption{\label{Fig.4}Squared dipole matrix element ($p_z$) between the ground bound subband $E_1$ and the continuum subbands $E_2$, $E_3$, $E_4$, $E_5$. The Dirac subbands are $E_3$ and $E_4$. The calculations are done for the SL satisfying $j = 1$, $j' = 2$ resonance condition. }
\end{figure}

We show in fig.\ref{Fig.4} the \textit{q} dependence of the modulus of the intersubband $p_z$ matrix element (from ground subband to higher energy subbands) for the 6.6 nm/21.37 nm SL satisfying the $j = 1$, $j' = 2$ resonance condition (see fig.\ref{Fig.2}(b) for the dispersion relation). 
%
In this superlattice there exists an almost dispersionless bound subband $E_1$ at about 34 meV. The first continuum subband $E_2$ is regular; hence the dispersions are parabolic in the vicinity of both $q = 0$ and $q = \pi/d$ and there is no degeneracy. Thus, at $q = 0$  ($q = \pi/d$) the superlattice wavefunctions should have the same (opposite) parities with respect to the centers of the layers \cite{Voisin1984}. As a result the $p_z$ intersubband matrix elements vanish at $q = 0$. This reasoning also applies to the other regular subbands. For the Dirac subbands with linear dispersions near $q = \pi/d$, we have found no such cancellations. Instead we find the same matrix elements at $q  = \pi/d$ as if subband 4 were the continuation of subband 3.

\begin{figure}
\includegraphics[scale=0.4]{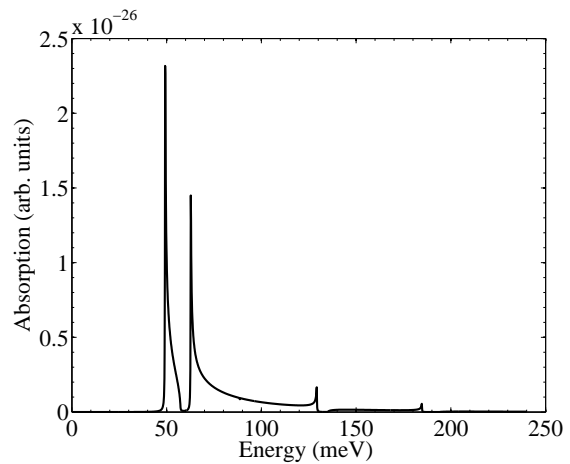}
\caption{\label{Fig.5}Absorption spectrum from ground subband towards higher energy subbands for the 6.6 nm /21.37 nm SL ($j = 1$, $j'= 2$).}
\end{figure}

The intersubband absorption lineshape for the 6.6 nm /21.37 nm SL is shown in fig.\ref{Fig.5}. 
%
The first peak corresponds to $E_1 \rightarrow E_2$ optical transitions around $q = \pi/d$.  The transition $E_1 \rightarrow E_2$ at $q = 0$ is parity forbidden and thus the associated absorption line is absent. The second peak corresponds to the $E_1 \rightarrow E_3$ transition at $q = 0$. It extends up to 129.1 meV which is the $E_1 \rightarrow E_4$ transition at $q = 0$. There is no hint of any feature around 149.2 meV which would correspond to the transitions $E_1 \rightarrow E_3$ and $E_1 \rightarrow E_4$ at $q = \pi/d$. Indeed, it can be readily checked that in the vicinity of this energy the absorption lineshape is a plateau (with the same amplitude before and after the critical energy). Finally, the $E_1 \rightarrow E_5$ transition starts smoothly at $q = 0$ (fig.\ref{Fig.4}) because it is parity forbidden at the zone center and ends up with a small singularity in the absorption spectrum at 184.7 meV, because the dipole matrix elements of $E_1 \rightarrow E_5$ is very small for any $q$.  A similar analysis could be made for the absorption spectrum of the superlattice satisfying the $j = j' =1$ resonance condition. 

\begin{figure}
\includegraphics[scale=0.9]{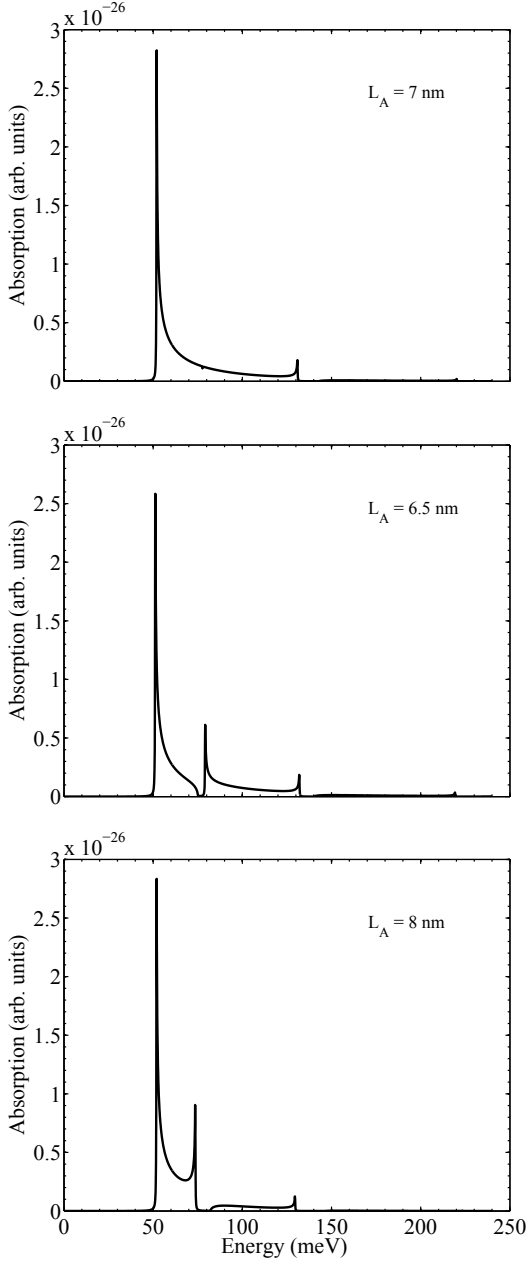}
\caption{\label{Fig.6}Comparison between the absorption coefficient of three SL's with the same period $d$ = 19.92 nm.  The Dirac SL corresponds to $j = j' = 1$ and $L_A$ = 7 nm.}
\end{figure}

It is interesting to compare what happens to the optical spectra when the layer thicknesses are changed slightly around those that realize a Dirac SL. Fig.\ref{Fig.6} shows the intersubband absorption for the Dirac SL with well thickness $L_A = 7$ nm and for two other SL's having the same period length 19.92 nm but well thickness of respectively 6.5 nm and 8 nm (where no resonance condition is satisfied). 
%
As shown in fig.\ref{Fig.2}(a) the Dirac SL satisfying the $j = j' = 1$ resonance condition has subbands $E_2$ and $E_3$ degenerate at $q = 0$ and located 77.8 meV above the ground subband. The three SL's share common optical features that are associated with the $E_1 \rightarrow E_2$ and $E_1 \rightarrow E_3$ optical absorption at $q = \pi/d$ (peaks at about 51 meV and 131 meV). Near 77 meV the SL's with $L_A$ = 6.5 nm and 8 nm show a transparency window. The 6.5 nm SL has a parity forbidden transition $E_1 \rightarrow E_2$ at $q = 0$ at the beginning of the transparency region while this $q = 0$ transition is allowed for the next absorption band (peak at 80 meV).  The reverse situation takes place for the SL with $L_A$ = 8 nm, the $q = 0$ optical transition being allowed (peak at 72.8 meV) then forbidden on each sides of the transparency region. The Dirac SL with $L_A$ = 7 nm resolves this parity change by showing no particular optical features (in particular no transparency region) at about 77 meV where the $E_1 \rightarrow E_2$ absorption ends and the $E_1 \rightarrow E_3$ absorption starts.

\begin{figure*}
\includegraphics[scale=0.9]{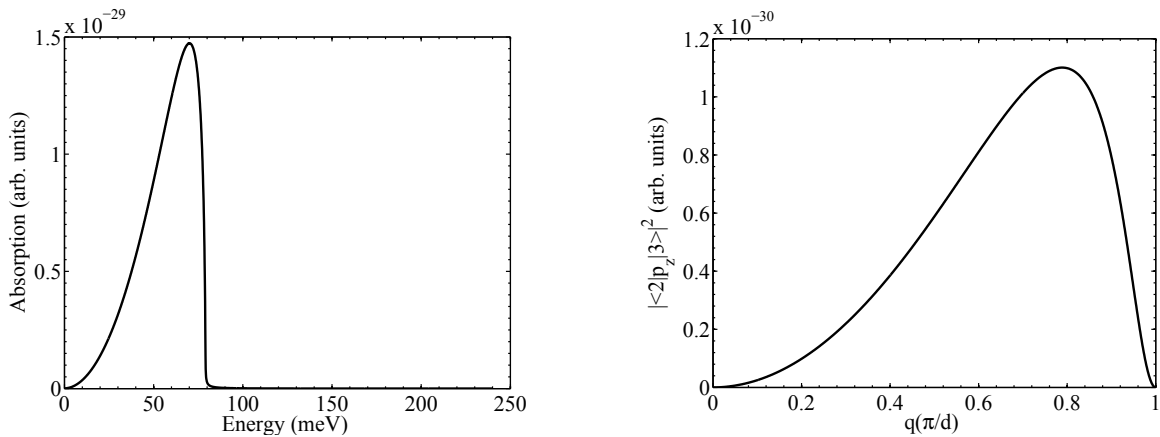}
\caption{\label{Fig.7}Absorption spectrum (left panel) and $p_z$ matrix element (right panel) for the optical transition from the lowest energy Dirac band to the higher energy one of the 7 nm /12.92 nm SL.}
\end{figure*}

Optical transitions between the Dirac subbands are allowed but weak as shown in the left panel of fig.\ref{Fig.7} for the $j = j' = 1$ SL. The corresponding $p_z$ matrix element is shown in the right panel of fig.\ref{Fig.7} and vanishes both at $q = 0$, because the degeneracy point is shown at the zone center, and at $q = \pi/d$ for parity reasons.
%
\section{Discussion}

\subsection{Stability against varying in-plane wavevector}
An interesting question is to examine whether the subband edge persists 
at non zero in-plane wavevector\citep{Bastard1991_SSP}.  Although the effective mass mismatch is small in the material we have chosen, the existence of degenerate bands either at $q = 0$ or $q = \pi/d$ may invalidate the usual perturbative treatments.
When there is a position dependent effective mass (piecewise constant), we need to find the eigenstates of the following Hamiltonian in presence of a magnetic field:
\begin{equation}
H=p_z\left(\frac{1}{2m(z)}\right) p_z+V_b(z)+\frac{1}{2m(z)}\left(p_{x}^{2}+(p_y+eBx)^{2}\right)
\label{eq.10}
\end{equation}
Stricto sensu, $H$ is not separable in $z$ and $(x,y)$.  However, one feels that if the effective masses are not too different this non separability is not so important.  Let us indeed split $H$ into a separable $H_0$ and a term supposed to be small.  We let:
\begin{equation}
\frac{1}{m(z)}=\Braket{\frac{1}{m}}+\left(\frac{1}{m(z)}-\Braket{\frac{1}{m}}\right)
\label{eq.11}
\end{equation}
where $\Braket{\frac{1}{m}}$ is not yet defined. The difference inside parentheses is expected to be small. Under such a circumstance we get:
%
\begin{eqnarray}
H=H_0+\delta H
\label{eq.12}
\end{eqnarray}
\begin{eqnarray}
H_0=p_z\left(\frac{1}{2m(z)}\right) p_z+V_b(z)+\frac{1}{2}\Braket{\frac{1}{m}}\left(p_{x}^{2}+(p_y+eBx)^{2}\right)
\nonumber
\end{eqnarray}
\begin{eqnarray}
\delta H=\frac{1}{2}\left(\frac{1}{m(z)}-\Braket{\frac{1}{m}}\right)\left(p_{x}^{2}+(p_y+eBx)^{2}\right)
\nonumber
\end{eqnarray}
The eigenstates and eigenvalues of $H_0$ are known: 
\begin{eqnarray}
\langle\overrightarrow{r}|{n,k_y,p,q}\rangle = \frac{1}{\sqrt{L_y}}\exp(ik_yy)\phi_n\left(x+\lambda ^2 k_y\right)\chi_{p,q}(z)
\nonumber
\end{eqnarray}
\begin{eqnarray}
\epsilon_{n,k_y,p,q}^0=\epsilon_p(q)+\left(n+1/2 \right)\hbar\langle \omega \rangle
\label{eq.13}
\end{eqnarray}
where $\langle \omega \rangle =eB\langle \frac{1}{m} \rangle$ and $\lambda=\sqrt{\frac{\hbar}{eB}}$. Now, in order to make the effects associated with $\delta H$ to be as small as possible, we impose that the first order correction to $\epsilon_{n,k_y,p,q}^0$ vanishes:
\begin{eqnarray}
\bra{n,k_y,p,q}\delta H|\ket{n,k_y,p,q}=
\label{eq.14}
\end{eqnarray}
\begin{eqnarray}
=\left(n+\frac{1}{2}\right)\hbar eB\bra{p,q}\left(\frac{1}{m(z)}-\Braket{\frac{1}{m}}\right)\ket{p,q}=
\nonumber
\end{eqnarray}
\begin{eqnarray}
=\left(n+\frac{1}{2}\right)\hbar \Braket{\omega}\left(\bra{p,q}\frac{1}{m(z)\Braket{\frac{1}{m}}}\ket{p,q}-1\right)=0
\nonumber
\end{eqnarray}
Hence, the unknown $\Braket{\frac{1}{m}}$ should be chosen such that $\Braket{\frac{1}{m}}=\bra{p,q}\frac{1}{m}\ket{p,q}$ if one wants the lack of separation between $z$ and $(x,y)$ motion to be minimized.  In practice, it is enough to ensure the equality in one elementary cell since the perturbation will collect all the cells' responses but also since the eigenstates are Bloch states.
Implicit in the previous reasoning is the non degeneracy of the state $\ket{p,q}$ . This is the case for most of the SL eigenstates except for the Dirac states. In the latter case we shall a priori define a $\Braket{\frac{1}{m}}=\frac{L_A}{m_A}+\frac{L_B}{m_B}$ and study the effect of $\delta H$ between two degenerate Dirac states. For the sake of definiteness, we shall study the effect of $\delta H$ at $q = \pi/d$ for a $j = 1$, $j'= 2$ resonance of the unperturbed Hamiltonian.
Due to the twofold degeneracy, there is some room to define the two eigenfunctions where to project $\delta H$.  These are:
\begin{eqnarray}
\chi^{(1)}_{\pi/d} = M \left \{
\begin{array}{r l}
-\sin\frac{\pi u}{L_A}  & -\frac{L_A}{2}\leq u \leq \frac{L_A}{2} \\
+\cos\frac{2\pi v}{L_B}  &  -\frac{L_B}{2}\leq v \leq \frac{L_B}{2}
\label{eq.15}
\end{array}
\right.
\end{eqnarray}
\begin{eqnarray}
\chi^{(2)}_{\pi/d} = M \left \{
\begin{array}{r l}
-\cos\frac{\pi u}{L_A}  & -\frac{L_A}{2}\leq u \leq \frac{L_A}{2} \\
+\xi\sin\frac{2\pi v}{L_B}  &  -\frac{L_B}{2}\leq v \leq \frac{L_B}{2}
\end{array}
\right.
\nonumber
\end{eqnarray}
where $u$ and $v$ refer to the position of the electron in layer $A$ and $B$ respectively and measured from the centers of the layers. $M$ and $N$ are constants obtained by normalizing the states in a SL period. With these wavefunctions one finds readily that $\delta H$ has the following matrix elements:
\begin{eqnarray}
\bra{\chi^{(1)}}\delta H \ket{\chi^{(1)}}=\bra{\chi^{(1)}}\delta H \ket{\chi^{(2)}}=0
\label{eq.16}
\end{eqnarray}
\begin{eqnarray}
\bra{\chi^{(2)}}\delta H \ket{\chi^{(2)}}=
\nonumber
\end{eqnarray}
\begin{eqnarray}
=\left(n+\frac{1}{2}\right)\frac{\hbar\Braket{\omega}}{\Braket{\frac{1}{m}}}\frac{L_AL_B}{L_A+L_B}\left(\frac{1}{m_A}-\frac{1}{m_B}\right)\frac{1-\xi^2}{L_A+\xi^2L_B}
\nonumber
\end{eqnarray}
with $m_A = 0.07m_0$, $m_B = 0.077m_0$, where $m_0$ is the free electron mass, $L_A = 6.6$ nm, $L_B = 21.3$ nm there is  $\xi = 1.752$ and $\bra{\chi^{(2)}}\delta H \ket{\chi^{(2)}}=A\left(n+\frac{1}{2}\right)\hbar\Braket{\omega}$ with $A = 1.41\times 10^{-2}$. At $B = 10$T and for  $n = 2$ there is:$\left(n+\frac{1}{2}\right)\hbar\Braket{\omega} =38.3$ meV. Hence the shift is -0.5 meV.  This value is indeed very small (actually much smaller than a typical broadening).  Thus, for the material parameters considered here the stability of Dirac feature against the in-plane wavevector is ensured.  Note that the conclusion may have to be reconsidered if the effective mass mismatch is more severe like in Ga(In)As/Al(In)As

\subsection{Impact on Bloch-oscillations}
The shape of the Dirac bands suggests in a semi-classical scenario of the Bloch oscillations ($\hbar dq/dt = -eF$, with $F$ the electric field) that Dirac bands should be associated with an  angular Bloch frequency of $eFd/2\hbar$. This is half the common value, as the carrier need to transverse two times the Brillioun zone, before the origin is reached again. However, it is not at all obvious that a semi-classical analysis applies to a situation where there is no gap between the two bands \cite{Bastard1994,Bouchard1995}. In order to observe Bloch oscillations, a sufficiently large electric field is needed, so that the Bloch frequency surpasses the scattering rate. This would lead to large Zener tunneling \cite{Glutsch2004} for the small gaps in the superlattices considered and thus makes the observation  difficult in actual semiconductor superlattices. Optical lattices \cite{Haller2010} with their absence of scattering may render the observation of Wannier quantization in Dirac SL's much easier.

\subsection{Dirac bands and inversion symmetry}
It is worth pointing out that the existence of Dirac bands in a binary SL is not related to the fact that the SL potential energy is centro-symmetric with respect to the center of one or the other layer that build the SL unit cell. Actually, we have found Dirac bands in the case of a polytype (ternary ABC superlattice) where the SL potential is non centro-symmetric. In quaternary superlattices, one may even find a Dirac band between the first two bands, as indicated by numerical findings in fig. 5 of Ref.\onlinecite{Romanova2011}. 

\section{Conclusion}
Previous works proved the existence of one dimensional gapless Dirac bands in semiconductor superlattices provided multiple Fabry-Perot conditions are fulfilled. In the present work we show that the dispersion relations close to the crossing point are linear. These Dirac SL's lay at the boundary of the SL parameters where the symmetry of the Wannier function changes. The existence of gapless Dirac bands implies interesting optical features that partly result from density of states considerations but more importantly reflect the change in the symmetry properties of the SL states. We also discussed the stability of the properties of Dirac SL's against varying in-plane wavevector.

\begin{acknowledgements}
We thank J. Dalibard and C. Sirtori for useful discussions.  The work at Lund University has been supported by the Swedish Research Council.
\end{acknowledgements}

\bibliographystyle{apsrev4-1}
\bibliography{mybiblio.bib}

\begin{thebibliography}{19}%
\makeatletter
\providecommand \@ifxundefined [1]{%
 \@ifx{#1\undefined}
}%
\providecommand \@ifnum [1]{%
 \ifnum #1\expandafter \@firstoftwo
 \else \expandafter \@secondoftwo
 \fi
}%
\providecommand \@ifx [1]{%
 \ifx #1\expandafter \@firstoftwo
 \else \expandafter \@secondoftwo
 \fi
}%
\providecommand \natexlab [1]{#1}%
\providecommand \enquote  [1]{``#1''}%
\providecommand \bibnamefont  [1]{#1}%
\providecommand \bibfnamefont [1]{#1}%
\providecommand \citenamefont [1]{#1}%
\providecommand \href@noop [0]{\@secondoftwo}%
\providecommand \href [0]{\begingroup \@sanitize@url \@href}%
\providecommand \@href[1]{\@@startlink{#1}\@@href}%
\providecommand \@@href[1]{\endgroup#1\@@endlink}%
\providecommand \@sanitize@url [0]{\catcode `\\12\catcode `\$12\catcode
  `\&12\catcode `\#12\catcode `\^12\catcode `\_12\catcode `\%12\relax}%
\providecommand \@@startlink[1]{}%
\providecommand \@@endlink[0]{}%
\providecommand \url  [0]{\begingroup\@sanitize@url \@url }%
\providecommand \@url [1]{\endgroup\@href {#1}{\urlprefix }}%
\providecommand \urlprefix  [0]{URL }%
\providecommand \Eprint [0]{\href }%
\providecommand \doibase [0]{http://dx.doi.org/}%
\providecommand \selectlanguage [0]{\@gobble}%
\providecommand \bibinfo  [0]{\@secondoftwo}%
\providecommand \bibfield  [0]{\@secondoftwo}%
\providecommand \translation [1]{[#1]}%
\providecommand \BibitemOpen [0]{}%
\providecommand \bibitemStop [0]{}%
\providecommand \bibitemNoStop [0]{.\EOS\space}%
\providecommand \EOS [0]{\spacefactor3000\relax}%
\providecommand \BibitemShut  [1]{\csname bibitem#1\endcsname}%
\let\auto@bib@innerbib\@empty
\bibitem [{\citenamefont {Bastard}(1988)}]{Bastard1988wave}%
  \BibitemOpen
  \bibfield  {author} {\bibinfo {author} {\bibfnamefont {G.}~\bibnamefont
  {Bastard}},\ }\href@noop {} {\emph {\bibinfo {title} {Wave mechanics applied
  to semiconductor heterostructures}}},\ Monographies de physique\ (\bibinfo
  {publisher} {Les {\'E}ditions de Physique},\ \bibinfo {year}
  {1988})\BibitemShut {NoStop}%
\bibitem [{\citenamefont {Sirtori}\ \emph {et~al.}(1994)\citenamefont
  {Sirtori}, \citenamefont {Capasso}, \citenamefont {Sivco},\ and\
  \citenamefont {Cho}}]{Sirtori1994}%
  \BibitemOpen
  \bibfield  {author} {\bibinfo {author} {\bibfnamefont {C.}~\bibnamefont
  {Sirtori}}, \bibinfo {author} {\bibfnamefont {F.}~\bibnamefont {Capasso}},
  \bibinfo {author} {\bibfnamefont {D.~L.}\ \bibnamefont {Sivco}}, \ and\
  \bibinfo {author} {\bibfnamefont {A.~Y.}\ \bibnamefont {Cho}},\ }\href@noop
  {} {\bibfield  {journal} {\bibinfo  {journal} {Applied Physics Letters}\
  }\textbf {\bibinfo {volume} {64}} (\bibinfo {year} {1994})}\BibitemShut
  {NoStop}%
\bibitem [{\citenamefont {Castro~Neto}\ \emph {et~al.}(2009)\citenamefont
  {Castro~Neto}, \citenamefont {Guinea}, \citenamefont {Peres}, \citenamefont
  {Novoselov},\ and\ \citenamefont {Geim}}]{Castro-Neto2009}%
  \BibitemOpen
  \bibfield  {author} {\bibinfo {author} {\bibfnamefont {A.~H.}\ \bibnamefont
  {Castro~Neto}}, \bibinfo {author} {\bibfnamefont {F.}~\bibnamefont {Guinea}},
  \bibinfo {author} {\bibfnamefont {N.~M.~R.}\ \bibnamefont {Peres}}, \bibinfo
  {author} {\bibfnamefont {K.~S.}\ \bibnamefont {Novoselov}}, \ and\ \bibinfo
  {author} {\bibfnamefont {A.~K.}\ \bibnamefont {Geim}},\ }\href {\doibase
  10.1103/RevModPhys.81.109} {\bibfield  {journal} {\bibinfo  {journal} {Rev.
  Mod. Phys.}\ }\textbf {\bibinfo {volume} {81}},\ \bibinfo {pages} {109}
  (\bibinfo {year} {2009})}\BibitemShut {NoStop}%
\bibitem [{\citenamefont {Orlita}\ and\ \citenamefont
  {Potemski}(2010)}]{Orlita2010}%
  \BibitemOpen
  \bibfield  {author} {\bibinfo {author} {\bibfnamefont {M.}~\bibnamefont
  {Orlita}}\ and\ \bibinfo {author} {\bibfnamefont {M.}~\bibnamefont
  {Potemski}},\ }\href {http://stacks.iop.org/0268-1242/25/i=6/a=063001}
  {\bibfield  {journal} {\bibinfo  {journal} {Semiconductor Science and
  Technology}\ }\textbf {\bibinfo {volume} {25}},\ \bibinfo {pages} {063001}
  (\bibinfo {year} {2010})}\BibitemShut {NoStop}%
\bibitem [{\citenamefont {Esaki}\ and\ \citenamefont {Tsu}(1970)}]{Esaki1970}%
  \BibitemOpen
  \bibfield  {author} {\bibinfo {author} {\bibfnamefont {L.}~\bibnamefont
  {Esaki}}\ and\ \bibinfo {author} {\bibfnamefont {R.}~\bibnamefont {Tsu}},\
  }\href {\doibase 10.1147/rd.141.0061} {\bibfield  {journal} {\bibinfo
  {journal} {IBM Journal of Research and Development}\ }\textbf {\bibinfo
  {volume} {14}},\ \bibinfo {pages} {61} (\bibinfo {year} {1970})}\BibitemShut
  {NoStop}%
\bibitem [{\citenamefont {Bastard}(1981)}]{Bastard1981}%
  \BibitemOpen
  \bibfield  {author} {\bibinfo {author} {\bibfnamefont {G.}~\bibnamefont
  {Bastard}},\ }\href {\doibase 10.1103/PhysRevB.24.5693} {\bibfield  {journal}
  {\bibinfo  {journal} {Phys. Rev. B}\ }\textbf {\bibinfo {volume} {24}},\
  \bibinfo {pages} {5693} (\bibinfo {year} {1981})}\BibitemShut {NoStop}%
\bibitem [{\citenamefont {Bastard}(1982)}]{Bastard1982}%
  \BibitemOpen
  \bibfield  {author} {\bibinfo {author} {\bibfnamefont {G.}~\bibnamefont
  {Bastard}},\ }\href {\doibase 10.1103/PhysRevB.25.7584} {\bibfield  {journal}
  {\bibinfo  {journal} {Phys. Rev. B}\ }\textbf {\bibinfo {volume} {25}},\
  \bibinfo {pages} {7584} (\bibinfo {year} {1982})}\BibitemShut {NoStop}%
\bibitem [{Note1()}]{Note1}%
  \BibitemOpen
  \bibinfo {note} {In the case of InAs/GaSb superlattices where the non
  parabolicity is a mandatory ingredient to understand the hybridization
  between the InAs electrons states and the GaSb light hole states, all the
  results discussed in terms of Fabry-Perot conditions as well as of linear
  dispersions apply. The only ingredient that changes in eq.\ref {eq.1} is the
  definition of the effective masses $m_A$ and $m_B$ which must account for non
  parabolicity effects \cite {Bastard1981, *Bastard1982}.}\BibitemShut {Stop}%
\bibitem [{\citenamefont {Bastard}\ \emph {et~al.}(1991)\citenamefont
  {Bastard}, \citenamefont {Brum},\ and\ \citenamefont
  {Ferreira}}]{Bastard1991_SSP}%
  \BibitemOpen
  \bibfield  {author} {\bibinfo {author} {\bibfnamefont {G.}~\bibnamefont
  {Bastard}}, \bibinfo {author} {\bibfnamefont {J.}~\bibnamefont {Brum}}, \
  and\ \bibinfo {author} {\bibfnamefont {R.}~\bibnamefont {Ferreira}},\ }in\
  \href {\doibase http://dx.doi.org/10.1016/S0081-1947(08)60092-2} {\emph
  {\bibinfo {booktitle} {Semiconductor Heterostructures and Nanostructures}}},\
  \bibinfo {series} {Solid State Physics}, Vol.~\bibinfo {volume} {44},\
  \bibinfo {editor} {edited by\ \bibinfo {editor} {\bibfnamefont
  {H.}~\bibnamefont {Ehrenreich}}\ and\ \bibinfo {editor} {\bibfnamefont
  {D.}~\bibnamefont {Turnbull}}}\ (\bibinfo  {publisher} {Academic Press},\
  \bibinfo {year} {1991})\ pp.\ \bibinfo {pages} {229 -- 415}\BibitemShut
  {NoStop}%
\bibitem [{\citenamefont {Ashcroft}\ and\ \citenamefont
  {Mermin}(1976)}]{Ashcroft1976}%
  \BibitemOpen
  \bibfield  {author} {\bibinfo {author} {\bibfnamefont {N.~W.}\ \bibnamefont
  {Ashcroft}}\ and\ \bibinfo {author} {\bibfnamefont {N.~D.}\ \bibnamefont
  {Mermin}},\ }\href@noop {} {\emph {\bibinfo {title} {Solid State Physics}}}\
  (\bibinfo  {publisher} {Holt, Rinehart and Winston},\ \bibinfo {address} {New
  York},\ \bibinfo {year} {1976})\BibitemShut {NoStop}%
\bibitem [{\citenamefont {Kohn}(1959)}]{Kohn1959_p1}%
  \BibitemOpen
  \bibfield  {author} {\bibinfo {author} {\bibfnamefont {W.}~\bibnamefont
  {Kohn}},\ }\href {\doibase 10.1103/PhysRev.115.809} {\bibfield  {journal}
  {\bibinfo  {journal} {Phys. Rev.}\ }\textbf {\bibinfo {volume} {115}},\
  \bibinfo {pages} {809} (\bibinfo {year} {1959})}\BibitemShut {NoStop}%
\bibitem [{\citenamefont {Bruno-Alfonso}\ and\ \citenamefont
  {Nacbar}(2007)}]{Bruno-Alfonso2007}%
  \BibitemOpen
  \bibfield  {author} {\bibinfo {author} {\bibfnamefont {A.}~\bibnamefont
  {Bruno-Alfonso}}\ and\ \bibinfo {author} {\bibfnamefont {D.~R.}\ \bibnamefont
  {Nacbar}},\ }\href {\doibase 10.1103/PhysRevB.75.115428} {\bibfield
  {journal} {\bibinfo  {journal} {Phys. Rev. B}\ }\textbf {\bibinfo {volume}
  {75}},\ \bibinfo {pages} {115428} (\bibinfo {year} {2007})}\BibitemShut
  {NoStop}%
\bibitem [{\citenamefont {Helm}\ \emph {et~al.}(1993)\citenamefont {Helm},
  \citenamefont {Hilber}, \citenamefont {Fromherz}, \citenamefont {Peeters},
  \citenamefont {Alavi},\ and\ \citenamefont {Pathak}}]{Helm1993}%
  \BibitemOpen
  \bibfield  {author} {\bibinfo {author} {\bibfnamefont {M.}~\bibnamefont
  {Helm}}, \bibinfo {author} {\bibfnamefont {W.}~\bibnamefont {Hilber}},
  \bibinfo {author} {\bibfnamefont {T.}~\bibnamefont {Fromherz}}, \bibinfo
  {author} {\bibfnamefont {F.~M.}\ \bibnamefont {Peeters}}, \bibinfo {author}
  {\bibfnamefont {K.}~\bibnamefont {Alavi}}, \ and\ \bibinfo {author}
  {\bibfnamefont {R.~N.}\ \bibnamefont {Pathak}},\ }\href {\doibase
  10.1103/PhysRevB.48.1601} {\bibfield  {journal} {\bibinfo  {journal} {Phys.
  Rev. B}\ }\textbf {\bibinfo {volume} {48}},\ \bibinfo {pages} {1601}
  (\bibinfo {year} {1993})}\BibitemShut {NoStop}%
\bibitem [{\citenamefont {Voisin}\ \emph {et~al.}(1984)\citenamefont {Voisin},
  \citenamefont {Bastard},\ and\ \citenamefont {Voos}}]{Voisin1984}%
  \BibitemOpen
  \bibfield  {author} {\bibinfo {author} {\bibfnamefont {P.}~\bibnamefont
  {Voisin}}, \bibinfo {author} {\bibfnamefont {G.}~\bibnamefont {Bastard}}, \
  and\ \bibinfo {author} {\bibfnamefont {M.}~\bibnamefont {Voos}},\ }\href
  {\doibase 10.1103/PhysRevB.29.935} {\bibfield  {journal} {\bibinfo  {journal}
  {Phys. Rev. B}\ }\textbf {\bibinfo {volume} {29}},\ \bibinfo {pages} {935}
  (\bibinfo {year} {1984})}\BibitemShut {NoStop}%
\bibitem [{\citenamefont {Bastard}\ \emph {et~al.}(1994)\citenamefont
  {Bastard}, \citenamefont {Ferreira}, \citenamefont {Chelles},\ and\
  \citenamefont {Voisin}}]{Bastard1994}%
  \BibitemOpen
  \bibfield  {author} {\bibinfo {author} {\bibfnamefont {G.}~\bibnamefont
  {Bastard}}, \bibinfo {author} {\bibfnamefont {R.}~\bibnamefont {Ferreira}},
  \bibinfo {author} {\bibfnamefont {S.}~\bibnamefont {Chelles}}, \ and\
  \bibinfo {author} {\bibfnamefont {P.}~\bibnamefont {Voisin}},\ }\href
  {\doibase 10.1103/PhysRevB.50.4445} {\bibfield  {journal} {\bibinfo
  {journal} {Phys. Rev. B}\ }\textbf {\bibinfo {volume} {50}},\ \bibinfo
  {pages} {4445} (\bibinfo {year} {1994})}\BibitemShut {NoStop}%
\bibitem [{\citenamefont {Bouchard}\ and\ \citenamefont
  {Luban}(1995)}]{Bouchard1995}%
  \BibitemOpen
  \bibfield  {author} {\bibinfo {author} {\bibfnamefont {A.~M.}\ \bibnamefont
  {Bouchard}}\ and\ \bibinfo {author} {\bibfnamefont {M.}~\bibnamefont
  {Luban}},\ }\href {\doibase 10.1103/PhysRevB.52.5105} {\bibfield  {journal}
  {\bibinfo  {journal} {Phys. Rev. B}\ }\textbf {\bibinfo {volume} {52}},\
  \bibinfo {pages} {5105} (\bibinfo {year} {1995})}\BibitemShut {NoStop}%
\bibitem [{\citenamefont {Glutsch}(2004)}]{Glutsch2004}%
  \BibitemOpen
  \bibfield  {author} {\bibinfo {author} {\bibfnamefont {S.}~\bibnamefont
  {Glutsch}},\ }\href {\doibase 10.1103/PhysRevB.69.235317} {\bibfield
  {journal} {\bibinfo  {journal} {Phys. Rev. B}\ }\textbf {\bibinfo {volume}
  {69}},\ \bibinfo {pages} {235317} (\bibinfo {year} {2004})}\BibitemShut
  {NoStop}%
\bibitem [{\citenamefont {Haller}\ \emph {et~al.}(2010)\citenamefont {Haller},
  \citenamefont {Hart}, \citenamefont {Mark}, \citenamefont {Danzl},
  \citenamefont {Reichs\"ollner},\ and\ \citenamefont {N\"agerl}}]{Haller2010}%
  \BibitemOpen
  \bibfield  {author} {\bibinfo {author} {\bibfnamefont {E.}~\bibnamefont
  {Haller}}, \bibinfo {author} {\bibfnamefont {R.}~\bibnamefont {Hart}},
  \bibinfo {author} {\bibfnamefont {M.~J.}\ \bibnamefont {Mark}}, \bibinfo
  {author} {\bibfnamefont {J.~G.}\ \bibnamefont {Danzl}}, \bibinfo {author}
  {\bibfnamefont {L.}~\bibnamefont {Reichs\"ollner}}, \ and\ \bibinfo {author}
  {\bibfnamefont {H.-C.}\ \bibnamefont {N\"agerl}},\ }\href {\doibase
  10.1103/PhysRevLett.104.200403} {\bibfield  {journal} {\bibinfo  {journal}
  {Phys. Rev. Lett.}\ }\textbf {\bibinfo {volume} {104}},\ \bibinfo {pages}
  {200403} (\bibinfo {year} {2010})}\BibitemShut {NoStop}%
\bibitem [{\citenamefont {Romanova}\ \emph {et~al.}(2011)\citenamefont
  {Romanova}, \citenamefont {Demidov}, \citenamefont {Mourokh},\ and\
  \citenamefont {Romanov}}]{Romanova2011}%
  \BibitemOpen
  \bibfield  {author} {\bibinfo {author} {\bibfnamefont {J.~Y.}\ \bibnamefont
  {Romanova}}, \bibinfo {author} {\bibfnamefont {E.~V.}\ \bibnamefont
  {Demidov}}, \bibinfo {author} {\bibfnamefont {L.~G.}\ \bibnamefont
  {Mourokh}}, \ and\ \bibinfo {author} {\bibfnamefont {Y.~A.}\ \bibnamefont
  {Romanov}},\ }\href {http://stacks.iop.org/0953-8984/23/i=30/a=305801}
  {\bibfield  {journal} {\bibinfo  {journal} {Journal of Physics: Condensed
  Matter}\ }\textbf {\bibinfo {volume} {23}},\ \bibinfo {pages} {305801}
  (\bibinfo {year} {2011})}\BibitemShut {NoStop}%
\end{thebibliography}%

\end{document}